\documentclass[prd,preprint,floatfix,nofootinbib,preprintnumbers,superscriptaddress,showkeys]{revtex4} 
\pdfoutput=1
\usepackage[caption=false]{subfig}
\usepackage{amsmath}
\usepackage{amsfonts}
\usepackage{amssymb}
\usepackage{float}
\usepackage{color}
\usepackage{graphicx}
\usepackage{graphics}
\usepackage{hyperref}
\usepackage{adjustbox,array}
\usepackage{enumitem} 
\hypersetup{}
\usepackage[utf8x]{inputenc} 
\usepackage{multirow}
\usepackage{booktabs}
\usepackage{mathrsfs}
\usepackage{diagbox}
\usepackage[english]{babel}
\usepackage[autostyle]{csquotes}

\graphicspath{{./figures/}}

\definecolor{rosso}{cmyk}{0,1,1,0.4}
\definecolor{rossos}{cmyk}{0,1,1,0.55}
\definecolor{rossoc}{cmyk}{0,1,1,0.2}
\definecolor{blu}{cmyk}{1,1,0,0.3}
\definecolor{blus}{cmyk}{1,1,0,0.6}
\definecolor{bluc}{cmyk}{1,1,0,0.1}
\definecolor{verde}{cmyk}{0.92,0,0.59,0.25}
\definecolor{verdec}{cmyk}{0.92,0,0.59,0.15}
\definecolor{verdes}{cmyk}{0.92,0,0.59,0.4}

\AtBeginDocument{
\heavyrulewidth=.08em
\lightrulewidth=.05em
\cmidrulewidth=.03em
\belowrulesep=.65ex
\belowbottomsep=0pt
\aboverulesep=.4ex
\abovetopsep=0pt
\cmidrulesep=\doublerulesep
\cmidrulekern=.5em
\defaultaddspace=.5em
}

\hypersetup{
 colorlinks=true,
 linkcolor=verdec,
 urlcolor=verdec,
 citecolor=verdec
}

\newcommand{\gsim}{\gtrsim}

\newcommand{\lf}{\left(}
\newcommand{\ri}{\right)}

\newcommand{\nn}{\nonumber}

\newcommand{\sqt}{\sqrt{2}}

\renewcommand{\lg}{\mathscr{L}} 

\newcommand{\br}{{\rm Br}}
\newcommand{\hc}{{\rm H.c.}}

\newcommand{\tot}{{\rm tot}}

\newcommand{\pb}{{\;{\rm pb}}}

\newcommand{\ifb}{{\;{\rm fb}^{-1}}}

\newcommand{\gev}{{\;{\rm GeV}}}

\newcommand{\beq}{\begin{equation}}
\newcommand{\eeq}{\end{equation}}
\newcommand{\bea}{\begin{eqnarray}}
\newcommand{\eea}{\end{eqnarray}}
\newcommand{\barr}{\begin{array}}
\newcommand{\earr}{\end{array}}
\newcommand{\bc}{\begin{center}}
\newcommand{\ec}{\end{center}}
\newcommand{\bit}{\begin{itemize}}
\newcommand{\eit}{\end{itemize}}
\newcommand{\ben}{\begin{enumerate}}
\newcommand{\een}{\end{enumerate}}

\newcommand{\al}{\alpha}
\newcommand{\bt}{\beta}

\newcommand{\sg}{\sigma}

\newcommand{\gm}{\gamma}
\newcommand{\Gm}{\Gamma}
\newcommand{\lm}{\lambda}

\newcommand{\hsm}{{h_{\rm SM}}}
\newcommand{\ch}{H^\pm}

\newcommand{\wpm}{W^\pm}
\newcommand{\wmp}{W^\mp}
\newcommand{\mh}{m_{h}}
\newcommand{\mch}{M_{H^\pm}}
\newcommand{\mhh}{M_{H}}
\newcommand{\ma}{M_{A}}
\newcommand{\mbsq}{\bar{m}^2}

\newcommand{\tb}{\tan\beta}
\newcommand{\sba}{\sin(\beta-\alpha)}
\newcommand{\cba}{\cos(\beta-\alpha)}

\newcommand{\mmu}      {{\mu^+ \mu^-}}

\newcommand{\ttop}      {{t\bar{t}}}

\definecolor{mint}{rgb}{0.24, 0.71, 0.54}

\begin{document}

\title{\color{verdes} 
Can a pseudoscalar with a mass of 365 GeV in two-Higgs-doublet models explain the CMS $t\bar{t}$ excess? }
\author{Chih-Ting Lu}
\email{ctlu@njnu.edu.cn}
\address{Department of Physics and Institute of Theoretical Physics, Nanjing Normal University, Nanjing, 210023, China}
\author{ Kingman Cheung}
\email{cheung@phys.nthu.edu.tw}
\address{Department of Physics, Konkuk University, Seoul 05029, Republic of Korea}
\address{Department of Physics, National Tsing Hwa University, Hsinchu 300, Taiwan}
\address{Center for Theory and Computation, National Tsing Hua University,
  Hsinchu 300, Taiwan}
\author{Dongjoo Kim}
\email{dongjookim.phys@gmail.com}
\address{Department of Physics, Konkuk University, Seoul 05029, Republic of Korea}
\author{Soojin Lee}
\email{soojinlee957@gmail.com}
\address{Department of Physics, Konkuk University, Seoul 05029, Republic of Korea}
\author{Jeonghyeon Song}
\email{jhsong@konkuk.ac.kr}
\address{Department of Physics, Konkuk University, Seoul 05029, Republic of Korea}

\begin{abstract}
We investigate the recently reported $\ttop$ excess by the CMS Collaboration within the framework of conventional Two-Higgs-Doublet Models (2HDMs). Considering all four types (I, II, X, and Y), we perform a comprehensive parameter space scan using the best-fit values for a pseudoscalar boson $A$: $M_A = 365$ GeV, $\Gamma_A/M_A = 2\%$, and $\tan\beta = 1.28$. Theoretical requirements and experimental constraints are systematically applied, including conditions from a bounded-below scalar potential, vacuum stability, unitarity, perturbativity, Flavor-Changing Neutral Currents (FCNCs), and direct searches at high-energy colliders. Our analysis shows that perturbativity imposes upper bounds of around 723 GeV on $M_{H^\pm}$ and $M_H$. FCNC constraints exclude all viable parameter space in Types II and Y, while a small region persists in Types I and X, but this region is ultimately ruled out by recent $t\bar{t} Z$ measurements by the  ATLAS and CMS Collaborations at the LHC. We conclude that conventional 2HDMs alone cannot accommodate a pseudoscalar boson that explains the observed $\ttop$ excess within viable parameter space. However, incorporating toponium effects in the background fit could potentially alter this conclusion.
\end{abstract}

\vspace{1cm}
\keywords{Higgs Physics, Beyond the Standard Model, electroweak precision data}

\maketitle
\tableofcontents

\section{Introduction}

Understanding the fundamental composition of matter and its interactions is a key objective in particle physics, with the Standard Model (SM) providing a solid foundation. Building upon this established framework, researchers explore potential new physics, and the Large Hadron Collider (LHC) plays a pivotal role in this endeavor. The LHC's dual mission involves conducting precision tests of SM observables~\cite{Erler:2019hds,Freitas:2020kcn} and searching for evidence of physics beyond the SM~\cite{Morrissey:2009tf,Golling:2016thc,Lyon:2024sdf}.

Given the LHC's status as a top quark factory~\cite{Han:2008xb,Bernreuther:2008ju,Cristinziani:2016vif}, the search for top quark pairs in final states is particularly significant. These searches enable precise measurements of top quark properties~\cite{ATLAS:2024dxp,Herwig:2019obz,ATLAS:2019onj,ATLAS:2019zrq,CMS:2019nrx,CMS:2022cqm,ATLAS:2023fsd,CMS:2024pts} and probe heavy resonances decaying into top quark pairs~\cite{ATLAS:2024vxm,ATLAS:2024jja}. However, measuring top quark pair ($t\overline{t}$) production near the threshold remains challenging due to complex non-perturbative QCD effects and intricate kinematic distributions~\cite{Sumino:1992ai,Hagiwara:2008df,Kiyo:2008bv,Sumino:2010bv,Ju:2020otc}.

Recently, the CMS experiment reported striking results showing an excess in the $t\bar{t}$ invariant mass near threshold at $\sqrt{s}=13$ TeV with a luminosity of 138 fb$^{-1}$~\cite{CMS-PAS-HIG-22-013}, significantly deviating from the perturbative QCD background-only hypothesis. The leading explanation, with  the highest difference in $-2\ln L$ and a significance exceeding $5\sigma$, is the existence of a spin-singlet $t\overline{t}$ bound state (toponium), 
$\eta_t$~\cite{Kuhn:1992ki,Kuhn:1992qw,Fabiano:1993vx,Beneke:1999qg,Kuhn:1987ty,Hagiwara:2008df,Kats:2009bv,Fuks:2021xje,Aguilar-Saavedra:2024mnm}. This interpretation suggests a production cross section of 7.1 pb with an uncertainty of $11\%$.
Another compelling explanation is a fundamental pseudoscalar $A$ with a mass of 365 GeV and a coupling to $t\overline{t}$ of $g_{Att}=0.78$, which fits the data well, albeit with a slightly lower $-2\ln L$ difference. Notably, the analysis of CP-sensitive observables strongly favors the pseudoscalar hypothesis over a scalar alternative. 

A follow-up theoretical study on the toponium hypothesis, including non-perturbative effects and its implications for the stability of the Higgs vacuum, has been conducted in Ref.~\cite{Matsuoka:2024pir}. Our study, however, focuses on the fundamental pseudoscalar explanation within the framework of two Higgs doublet models (2HDMs)~\cite{Branco:2011iw,BhupalDev:2014bir,Bhattacharyya:2015nca,Darvishi:2019ltl,Wang:2022yhm,Darvishi:2023fjh}, which were discussed as a prototype for the single $A$ hypothesis in the CMS study~\cite{CMS-PAS-HIG-22-013}.

The best-fit values for a fundamental pseudoscalar interpretation ($M_A = 365$ GeV, $\Gamma_A/M_A = 2\%$, and $g_{Att} = 0.78$) can be naturally interpreted within conventional 2HDMs. However, it is crucial to verify whether these parameters, while explaining the observed excess, are consistent with the broader theoretical and experimental constraints. To this end, we conduct a detailed investigation into the viability of this pseudoscalar $A$ within the 2HDM framework, considering theoretical requirements~\cite{Eriksson:2009ws}, Flavor-Changing Neutral Currents (FCNC)~\cite{Misiak:2017bgg,Arbey:2017gmh,Sanyal:2019xcp,HFLAV:2019otj}, electroweak precision data (EWPD)~\cite{ParticleDataGroup:2024cfk}, and relevant collider constraints~\cite{Bahl:2022igd}, especially recent $t\overline{t}Z$ search results at the LHC~\cite{ATLAS:2023szc,CMS:2024mtn}. Our exploration aims to provide significant implications for future model building and experimental searches in the context of this observed excess.

This paper is organized as follows: In Section \ref{sec:review}, we review the essential features of 2HDMs relevant to our analysis. Section \ref{sec:CMStt} discusses how the 2HDM framework can naturally explain the best-fit values associated with the CMS $t\bar{t}$ excess. In Section \ref{sec:viability}, we present our results, discussing the implications of each set of constraints on the viability of the 2HDM interpretation. Finally, we conclude in Section \ref{sec:conclusions}, summarizing our findings and their implications for future searches and model building in the context of the observed $t\bar{t}$ excess.

\section{Review of 2HDM}
\label{sec:review}

The 2HDM introduces two complex $SU(2)_L$ Higgs doublet fields, $\Phi_1$ and $\Phi_2$~\cite{Branco:2011iw}:
\bea
\label{eq-phi:fields}
\Phi_i = \left( \begin{array}{c} w_i^+ \\[3pt]
\dfrac{v_i +  h_i + i \eta_i }{ \sqrt{2}}
\end{array} \right), \quad i=1,2,
\eea
where $v_{1}$ and $v_2$ are the nonzero vacuum expectation values of $\Phi_1$ and $\Phi_2$, respectively.
Electroweak symmetry breaking occurs at $v =\sqrt{v_1^2+v_2^2}=246\gev$, with $\tan \beta =v_2/v_1$.

To avoid tree-level FCNC~\cite{Glashow:1976nt,Paschos:1976ay}, we impose a discrete $Z_2$ symmetry
($\Phi_1 \to \Phi_1$, $\Phi_2 \to -\Phi_2$).
The $CP$-invariant scalar potential with softly broken $Z_2$ symmetry is:
\bea
\label{eq-VH}
V = && m^2 _{11} \Phi^\dagger _1 \Phi_1 + m^2 _{22} \Phi^\dagger _2 \Phi_2
-m^2 _{12} ( \Phi^\dagger _1 \Phi_2 + \hc) \\ \nn
&& + \frac{1}{2}\lambda_1 (\Phi^\dagger _1 \Phi_1)^2
+ \frac{1}{2}\lambda_2 (\Phi^\dagger _2 \Phi_2 )^2
+ \lambda_3 (\Phi^\dagger _1 \Phi_1) (\Phi^\dagger _2 \Phi_2)
+ \lambda_4 (\Phi^\dagger_1 \Phi_2 ) (\Phi^\dagger _2 \Phi_1) \\ \nn
&& + \frac{1}{2} \lambda_5
\left[
(\Phi^\dagger _1 \Phi_2 )^2 +  \hc
\right],
\eea
where the $m^2 _{12}$ term softly breaks the $Z_2$ parity.
The model yields five physical Higgs bosons: $h$ (lighter $CP$-even), 
$H$ (heavier $CP$-even), $A$ ($CP$-odd), and $H^\pm$ (charged). 
Weak eigenstates are related to physical states via mixing angles $\al$ and $\bt$~\cite{Song:2019aav}. 

The SM Higgs boson $\hsm$ is related to $h$ and $H$ by
\bea
\label{eq-hsm}
\hsm = \sba h + \cba H.
\eea
For SM-like Higgs behavior~\cite{ATLAS:2022vkf,CMS:2022dwd}, 
we adopt the Higgs alignment limit: $h$ as the observed Higgs ($\mh=125\gev$, $\sba=1$)~\cite{Carena:2013ooa,Celis:2013rcs,Cheung:2013rva,Bernon:2015qea,Chang:2015goa,Das:2015mwa,Kanemura:2021dez}. This prohibits $H \to WW/ZZ$, $A \to Z h$, and $\ch\to W^{\pm (*)} h$ at tree level.

The quartic couplings are given by~\cite{Das:2015mwa}
\begin{align}
\label{eq-lm1}
\lm_1 &= \frac{1}{v^2 }
\left[
\tb^2 (\mhh^2 - \mbsq) -  m_h^2 
\right], \\[3pt] \label{eq-lm2}
\lm_2 &= \frac{1}{v^2 }
\left[ \mh^2 + \frac{1}{\tb^2} \lf \mhh^2-\mbsq \ri
\right], \\[3pt] \label{eq-lm3}
\lm_3&= 
\frac{1}{v^2}
\left[ \mh^2 +2 \mch^2 -\mhh^2-\mbsq
\right], \\[3pt] \label{eq-lm4}
\lm_4 &=
\frac{1}{v^2}
\left[
\ma^2- 2 \mch^2 + \mbsq
\right], \\[3pt] \label{eq-lm5}
\lm_5 &=
\frac{1}{v^2}
\left[
\mbsq-\ma^2
\right], 
\end{align}
where $\mbsq = m_{12}^2/(\sin\beta\cos\beta)$.

The Yukawa couplings of the SM fermions to the Higgs bosons depend 
on the $Z_2$ parity of the fermion singlets, yielding four types in the 2HDM: 
Type-I, Type-II, Type-X, and Type-Y. 
The Yukawa Lagrangian is parametrized as:
\bea
\label{eq-Lg:Yukawa}
\lg_{\rm Yuk} &=&
- \sum_f 
\lf 
\frac{m_f}{v}  \bar{f} f h + \frac{m_f}{v} \xi^H_f \bar{f} f H
-i \frac{m_f}{v} \xi^A_f \bar{f} \gm_5 f A
\ri
\\ \nn &&
- 
\left\{
\dfrac{\sqrt2V_{ud}}{v } H^+  \overline{u}
\left(m_u \xi^A_u \text{P}_L +  m_d \xi^A_d \text{P}_R\right)d 
+\dfrac{\sqt m_\ell}{v}H^+ \xi^A_\ell \overline{\nu}_L\ell_R^{}
+\hc
\right\},
\eea
with Yukawa coupling modifiers in the alignment limit:
\begin{align}
\text{Type-I: } & \xi^H_u = \xi^H_d =\xi^H_\ell = \xi^A_u = -\xi^A_d = - \xi^A_\ell=\frac{1}{\tb},
\\ \nn
\text{Type-II: } & \xi^H_u =  \xi^A_u =\frac{1}{\tb}, \quad 
\xi^H_d =\xi^H_\ell =\xi^A_d =  \xi^A_\ell=\tb.
\end{align}
Through our comprehensive analysis, we found that the viability of Type-I (Type-II) in explaining the CMS $t\bar{t}$ excess is nearly identical to that of Type-X (Type-Y). Therefore, we focus on presenting the results for Type-I and Type-II.

\section{CMS $t\bar{t}$ Excess and 2HDM Interpretation }
\label{sec:CMStt}

\begin{figure}[t] \centering
\begin{center}
\includegraphics[width=0.85\textwidth]{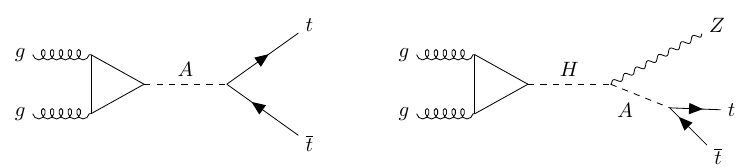}
\end{center}
\caption{\label{fig-Feynman}
Feynman diagrams for $gg\to A \to \ttop$ (left panel) and for $gg\to H \to Z A(\to \ttop)$ (right panel).}
\end{figure}

The CMS collaboration, using 138 $\ifb$ of data at the 13 TeV LHC, reported an excess in the $t\bar{t}$ invariant mass spectrum around 365 GeV. They analyzed three interpretations: a $^1S_0^{(1)}$ $t\bar{t}$ bound state $\eta_t$, a single pseudoscalar $A$, and a single heavy CP-even neutral scalar $H$. The Feynman diagram for the pseudoscalar $A$ hypothesis, which involves gluon fusion production followed by the decay $A \to t\bar{t}$, is shown in the left panel of \autoref{fig-Feynman}.

The CMS presented the values of $\Delta \left( -2 \ln L\right)$, where $L$ is the likelihood, 
between the best-fit point for each hypothesis and the background-only hypothesis:
\begin{align}
\label{eq-CMS}
\eta_t &: & m_{\eta_t}&=343\gev,  &  \Gm_{\eta_t} &= 7\gev, & & & \Delta \left( -2 \ln L\right)&= -86.2,
\\ \nn
A &: & \ma &= 365\gev ,  &  \Gm_{A} &= 0.02 \ma,&  \tb&=1.28,&  \Delta \left( -2 \ln L\right)&= -72.6,
\\ \nn
H &: & \mhh &= 365\gev ,  &  \Gm_{H} &= 0.02 \mhh, & \tb&=0.69,&  \Delta \left( -2 \ln L\right)&= -10.4.
\end{align}
The excess is most compatible with the $\eta_t$ hypothesis, with a global significance exceeding five standard deviations and an excess cross section of $7.1 \pm 0.8$ pb.

The single pseudoscalar $A$ hypothesis emerges as the second strongest candidate, 
with CMS reporting that its local significance exceeds $5\sigma$ deviations 
from the background-only hypothesis.\footnote{To explain similar LHC excesses in ditop and ditau channels,
a pseudoscalar resonance with a mass of 400 GeV was studied in Ref.~\cite{Arganda:2021yms}.}
The analysis significantly favors the pseudoscalar over the scalar hypothesis, 
a preference supported by two CP-sensitive observables, $c_{\rm hel}$~\cite{Bernreuther:2004jv,Bernreuther:2015yna,Aguilar-Saavedra:2022uye} and $c_{\rm han}$~\cite{Aguilar-Saavedra:2022uye}. 

To compare the $\eta_t$ and $A$ hypotheses in explaining the CMS $t\bar{t}$ excess, we estimate the global significance of the $A$ hypothesis. It is reported that the $\eta_t$ hypothesis achieves a $5\sigma$ significance with $\Delta \left( -2 \ln L\right) = -86.2$. Since $\eta_t$ has one free parameter ($\Gm_{\eta_t}$), this implies 32 data points. Using the $\Delta \left( -2 \ln L \right)$ for the $A$ hypothesis, we infer the degrees of freedom, noting that the $A$ hypothesis involves four model parameters: $\tb$, $\ma$, $\mch$, and $\mhh$. This yields a global significance of $4.8\sigma$, providing strong support for the $A$ hypothesis.

However, a straightforward calculation of the cross section times branching ratio for the $A$ hypothesis reveals a notable discrepancy with the observed excess. At the best-fit point ($\ma=365\gev$ and $\tb=1.28$), the Next-to-Next-to-Next-to-Leading Logarithm (N$^3$LL) prediction for the cross section is $\sigma(gg \to A) \approx 26.2\pb$~\cite{Ahmed:2015qda,Ahmed:2016otz}. Even assuming $\br(A \to t\bar{t}) = 1$, this value significantly overshoots the reported excess cross section of approximately 7.1 pb.

This discrepancy highlights two key factors. First, there is significant destructive interference between the $gg \to A \to t\bar{t}$ process and the QCD background $gg \to t\bar{t}$, which reduces the observed cross section. Notably, this interference is an inherent feature of the 2HDM framework. Second, for the $A$ hypothesis to match the observed excess, the branching ratio $\br(A \to t\bar{t})$ must indeed be very close to 1.

\begin{figure}[t] \centering
\begin{center}
\includegraphics[width=0.85\textwidth]{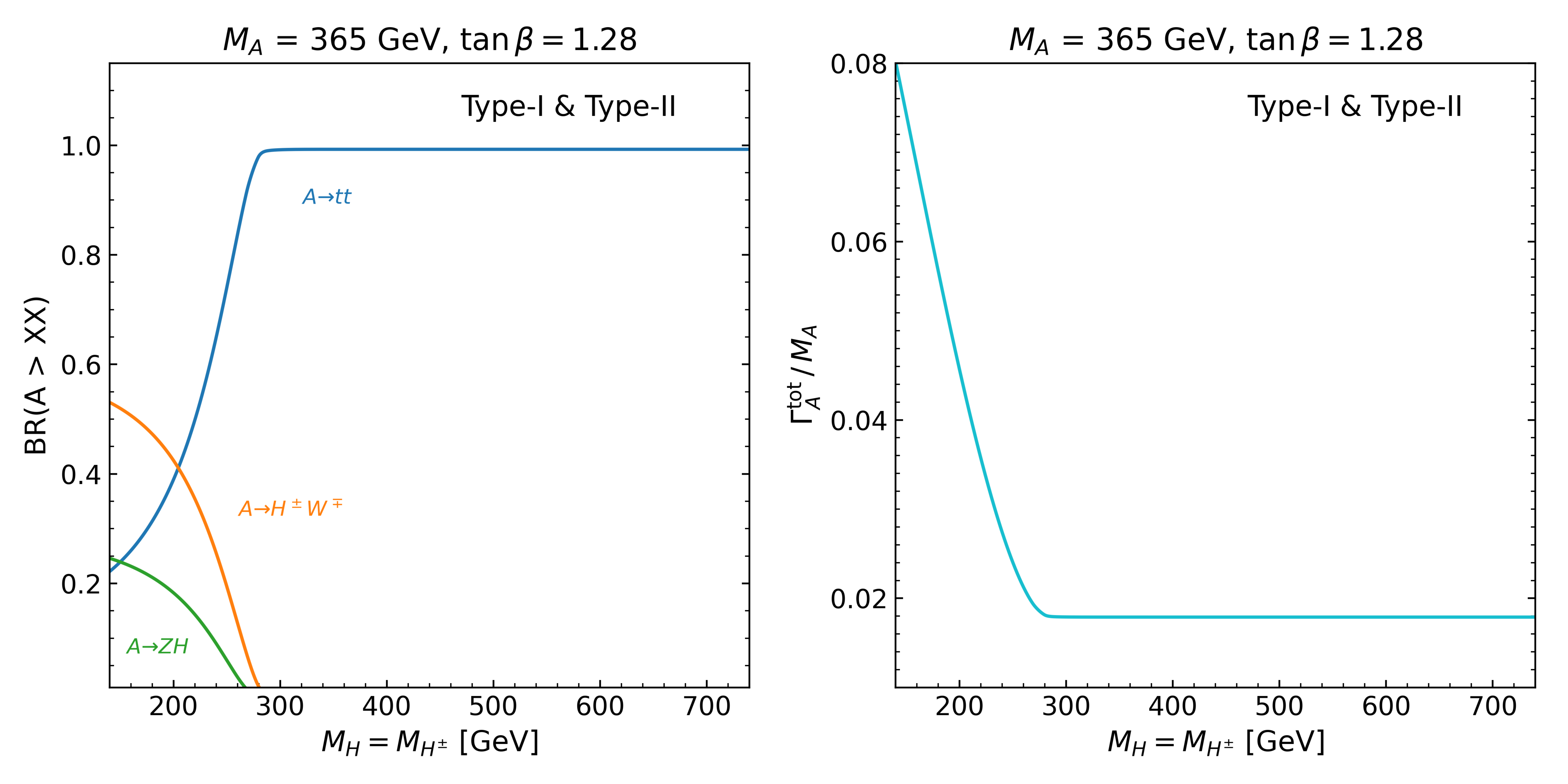}
\end{center}
\caption{\label{fig-BRA-GmA}
Branching ratios for the pseudoscalar $A$ (left panel) and the ratio of the total decay width to $\ma$ (right panel) as a function of $\mch$. 
We set $\ma=365\gev$ and $\tb=1.28$. Solid and dashed lines represent Type-I and Type-II results, respectively, which are nearly indistinguishable due to their close similarity. }
\end{figure}

Based on the CMS best-fit results and our analysis of the $A$ hypothesis, 
we consider the following 2HDM setup:
\bea
\label{eq-setup}
\mh=125\gev,\quad \sba=1,\quad \ma=365\gev, \quad \tb=1.28 \pm 0.128.
\eea
Note that we allow a 10\% uncertainty in the value of $\tb$.

To assess the feasibility of achieving $\br(A\to \ttop)\simeq 1$ and $\Gm_A/\ma \simeq 2\%$,
we present in \autoref{fig-BRA-GmA} the branching ratios for the pseudoscalar $A$ (left panel)
and the ratio of its total decay width to $\ma$ (right panel) as a function of $\mch$.
We omit decay modes with branching ratios below 1\%, such as $\br(A\to b\bar{b})$.
For simplicity, we assume $\mch=\mhh$ in this figure.
The results are calculated for $\ma=365\gev$ and $\tb=1.28$. Notably, the outcomes for Type-I and Type-II 2HDMs, represented by solid and dashed lines respectively, are virtually identical, reflecting the similarity of these models under the given parameters.

The branching ratios of $A$ show a strong dependence on the masses of $H$ and $H^\pm$. 
For lighter $\mch(=\mhh)$, the decays $A\to \ch\wmp$ and $A\to Z H$ become substantial, 
reducing $\br(A\to\ttop)$. 
Conversely, when $\ch$ and $H$ have masses exceeding about 280 GeV, we achieve
the required condition of $\br(A\to\ttop)\simeq 1$. 

The right panel of \autoref{fig-BRA-GmA} illustrates $\Gm^\tot_A/\ma$ as a function of $\mch$. 
For $H^\pm$ and $H$ masses below about 280 GeV, 
the sizable partial widths of $A\to \ch\wmp$ and $A\to Z H$ enhance $\Gm^\tot_A$, 
resulting in $\Gm^\tot_A/\ma$ significantly larger than 2\%. 
In contrast, heavier $H^\pm$ and $H$ (above $\sim 280 \gev$) reduce $\Gm^\tot_A/\ma$ 
to approximately 1.8\%, remarkably close to the CMS best-fit value.
Therefore, to explain the CMS $\ttop$ excess cross section, both $\mch$ and $\mhh$ should be greater than approximately 280 GeV.

In conclusion, these results demonstrate that the pseudoscalar $A$ in the 2HDM 
can naturally accommodate the observed characteristics of the CMS $\ttop$ excess, 
providing strong motivation for further investigation of this scenario.

\section{Viability of 2HDM Scenarios Explaining the CMS $\ttop$ Excess}
\label{sec:viability}

Having shown that the pseudoscalar $A$ in the 2HDM can account for the best-fit point of the CMS $\ttop$ excess, we now examine whether the parameter points explaining this excess also satisfy key theoretical and experimental constraints. To achieve this, we randomly scan the model parameters and cumulatively impose theoretical and experimental constraints.

For the setup in \autoref{eq-setup}, we scan the parameter ranges of
\begin{align}
\label{eq-scan-range}
\tb &\in [1.152,1.408],\quad m_{12}^2\in \left[ 0, 1000^2 \right] \gev^2, 
\\ \nn
\mhh &\in \left[ 130, 1500 \right] \gev, \quad \mch \in \left[ 200, 1500 \right] \gev.
\end{align}
We randomly generate uniformly distributed four-dimensional parameter points.

We cumulatively impose the following constraints:
\begin{description}
\item[Step-(i) Theory0+EWPD:]
We require parameter points to satisfy three theoretical requirements and the Peskin-Takeuchi oblique parameters $S$ and $T$~\cite{He:2001tp,Grimus:2008nb,ParticleDataGroup:2024cfk}. 
The theoretical conditions include bounded-from-below Higgs potential~\cite{Ivanov:2006yq}, 
tree-level unitarity of scalar-scalar, scalar-vector, and vector-vector scattering amplitudes~\cite{Kanemura:1993hm,Akeroyd:2000wc}, 
and vacuum stability~\cite{Barroso:2013awa}. For the oblique parameters, we use the 2024 Particle Data Group results~\cite{ParticleDataGroup:2024cfk}:
\begin{align}
\label{eq-STU}
S &= - 0.04 \pm 0.10 , \quad T = 0.01 \pm 0.12 , \quad U = - 0.01 \pm 0.09 ,
\\ \nn
\rho_{ST} &= 0.93, \quad
\rho_{SU} = -0.70, \quad
\rho_{TU} = -0.87,
\end{align}
where $\rho_{ij}$ are the correlation coefficients between parameters.

\item[Step-(ii) Perturbativity:]
We require that quartic Higgs couplings among physical Higgs bosons satisfy $\left| C_{H_i H_j H_k H_l}\right|<4\pi$~\cite{Branco:2011iw,Chang:2015goa}.

\item[Step-(iii) FCNC:]
We impose constraints from flavor physics observables, primarily $b\to s \gm$ and $B_d \to \mmu$~\cite{Haller:2018nnx,Arbey:2017gmh,Misiak:2017bgg,HFLAV:2019otj,Misiak:2020vlo}.
	
\item[Step-(iv) Collider:] We apply collider constraints, including Higgs precision data and direct search bounds from LEP, Tevatron, and LHC, using \textsc{HiggsTools}~\cite{Bahl:2022igd}.
\end{description}
For Steps (i) and (ii), we employ 2HDMC version 1.8.0~\cite{Eriksson:2010zzb}.

\begin{figure}[t] \centering
\begin{center}
\includegraphics[width=0.95\textwidth]{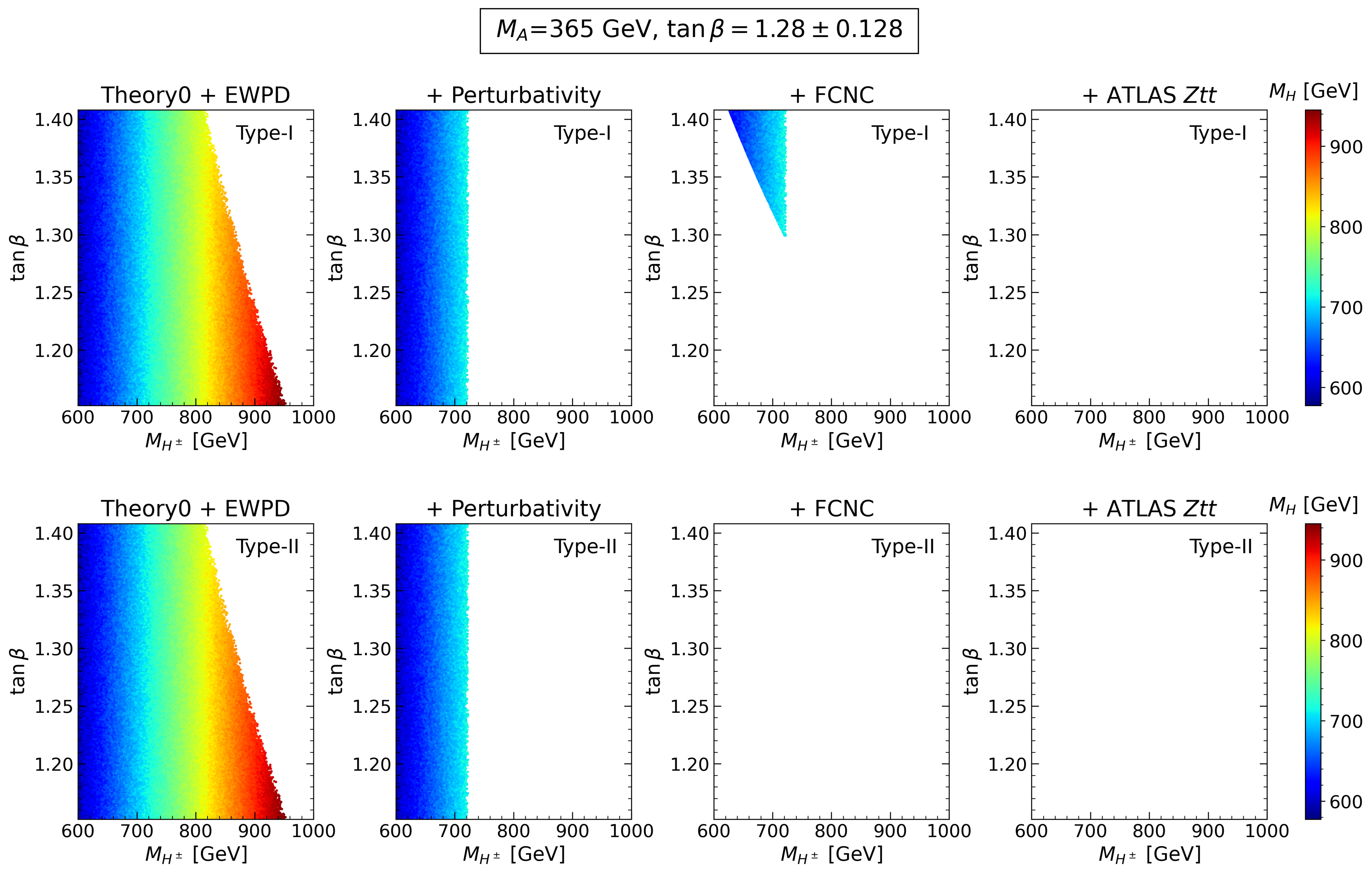}
\end{center}
\caption{\label{fig-survived-param}
Viable parameter space in the ($M_{H^\pm}$, tan$\beta$) plane, with $M_H$ indicated by the color scale. The upper and lower rows show results for Type-I and Type-II 2HDMs, respectively. Each column represents the parameter space after applying cumulative constraints:
(1) bounded-from-below Higgs potential, vacuum stability, unitarity, and oblique parameters;
(2) Perturbativity requirements;
(3) Flavor physics constraints;
(4) Higgs precision data and direct LHC search bounds.
All panels assume $\ma= 365\gev$ and $\tb = 1.28  \pm 0.128$.  }
\end{figure}

For both Type-I and Type-II 2HDMs, we generated $10^6$ parameter points each satisfying the criteria in Step (i). The survived parameter points after each step are presented in \autoref{fig-survived-param} for Type-I (upper panels) and Type-I (lower panels). Each column represents the parameter space after applying cumulative constraints from Steps (i) to (iv). All panels assume $M_A = 365$ GeV and $\tb = 1.28 \pm 0.128$.

The allowed parameter points after Step (i) show clear characteristics. 
First, there exist upper bounds on $\mch$ and $\mhh$ below about 950 GeV, 
mainly due to unitarity constraints on the quartic couplings $\lm_{1,\cdots,5}$ in \autoref{eq-VH}. 
Given $\lm_4-\lm_5 = 2\left(\ma^2-\mch^2\right)/v^2$ from \autoref{eq-lm4} and \autoref{eq-lm5}, 
the fixed $\ma=365\gev$ restricts $\mch$. 
Secondly, there is a strong correlation between $\mch$ and $\mhh$, 
indicated by the vertical color gradient, 
suggesting similar masses for these two states. 
This feature arises from oblique parameter constraints, 
which limit the mass difference between $A$ or $H$ and the charged Higgs boson~\cite{Lee:2022gyf}.

The panels in the second column display the viable parameter points that satisfy the perturbativity condition. The impact of this requirement is both significant and extensive, excluding the vast majority of the parameter space that survived Step (i). In particular, all parameter points with a charged Higgs mass (and consequently, the heavier CP-even neutral Higgs boson) greater than approximately 723 GeV are ruled out. This substantial reduction in viable parameter space highlights the tension between maintaining perturbativity and the need for relatively heavy additional Higgs bosons to evade direct search bounds.

While violations of perturbativity can sometimes be tolerated in theoretical models, our detailed analysis uncovers a critical issue unique to this scenario. These violations predominantly affect couplings involving the observed Higgs boson, specifically $C_{h H_i H_j H_k}$. This is especially problematic because such violations would result in significant deviations from the SM-like behavior of the Higgs boson at loop level, rendering them incompatible with current experimental data. Therefore, in this case, we cannot relax the perturbativity condition without compromising the model's consistency with observations of the 125 GeV Higgs boson.

The third column panels incorporate FCNC observables, which play a crucial role. In Type-I, only a small triangular region in the $(\mch, \tb)$ parameter space remains viable. For Type-I, the most stringent constraint comes from $B\to \mmu$~\cite{Haller:2018nnx}, which requires $\mch > 740\gev$ at $\tb = 1.28$. As the lower bound on $\mch$ gradually decreases with increasing $\tb$, only parameter points with $\tb > 1.3$ survive at Step (iii). In contrast, no region survives in Type-II. The most significant constraint comes from $b \to s \gamma$~\cite{Misiak:2020vlo}, which requires $\mch > 800\gev$ for $\tb \gsim 0.8$. Since perturbativity already imposes an upper bound of $\mch < 723\gev$, no parameter points in Type-II can satisfy both conditions simultaneously.

The fourth column panels show the remaining parameter points after imposing constraints from Higgs precision data and direct search bounds. At this stage, all surviving points for Type-I are excluded. Our detailed investigation reveals that a single process, $pp \to Z t\bar{t}$~\cite{ATLAS:2023szc, CMS:2024mtn}, eliminates the remaining parameter points in Type-I after Step (iii). Although this search specifically targeted a CP-odd Higgs boson decaying into a heavy CP-even Higgs boson and a $Z$ boson in the $\ell^+\ell^- t\bar{t}$ and $\nu \bar{\nu} t\bar{t}$ final states, the null results do not uniquely determine the CP nature of the $t\bar{t}$ parent particle. Thus, these results should also be applied to the process $pp \to H \to Z A$, followed by $A \to t\bar{t}$ (as shown in the right panel of \autoref{fig-Feynman}). Since the upper bound on $\mhh$ after Step (iii) is $\mhh < 723\gev$, which is low enough to produce a large cross section for $gg \to H$, the $Z t\bar{t}$ constraint ultimately excludes Type-I.

\begin{figure}[t] \centering
\begin{center}
\includegraphics[width=0.85\textwidth]{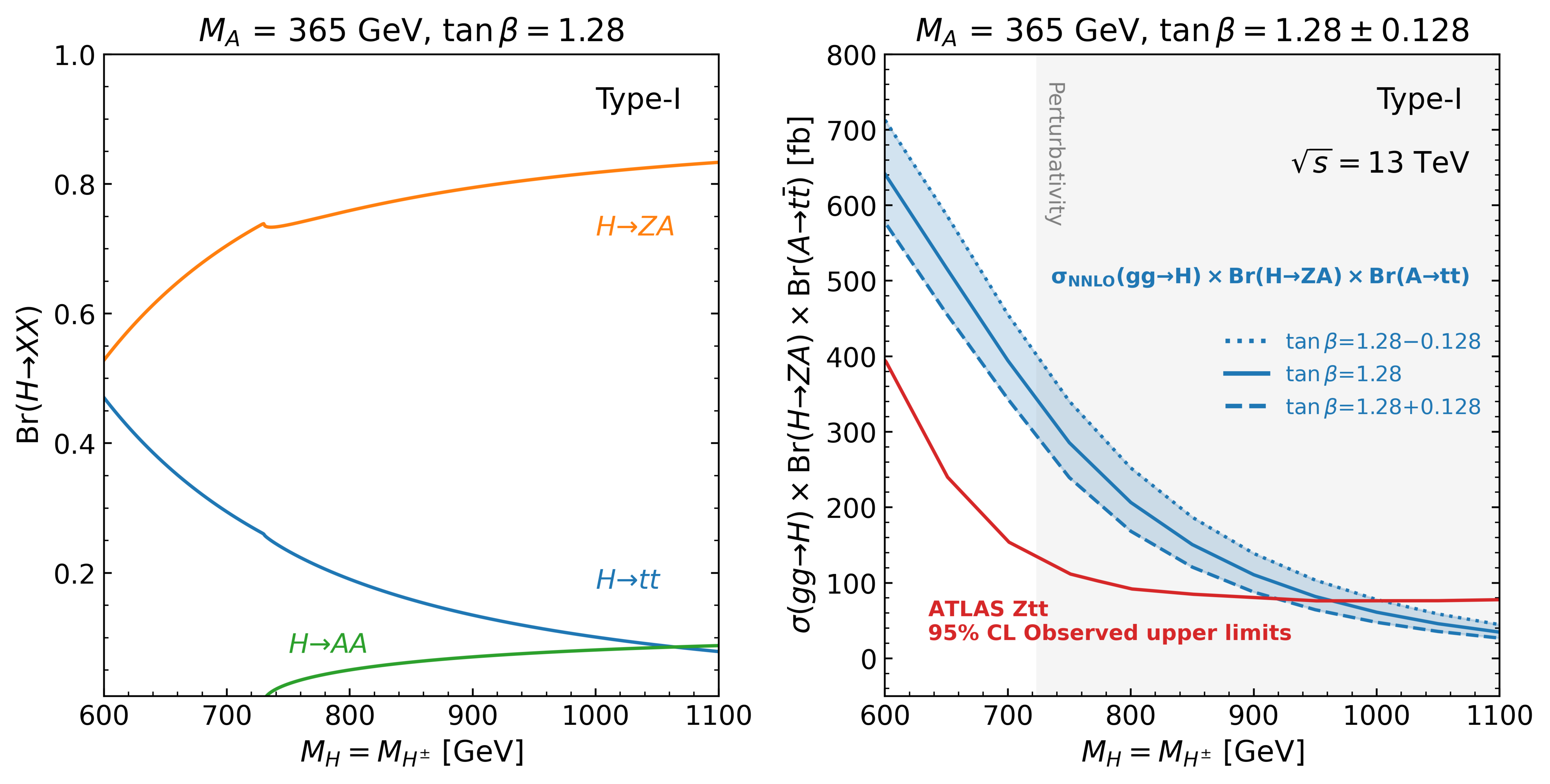}
\end{center}
\caption{\label{fig-BRH-Ztt}
Branching ratios of $H$ as a function of $\mhh$ (left panel)
and $\sg_{\rm NNLO}(gg\to H)\br(H \to ZA) \br(A\to\ttop)$ at the 13 TeV LHC (right panel).
We set $\ma=365\gev$, $\tb=1.28$, $m_{12}^2 = 7\times 10^4 \gev^2$,
and $\mch=\mhh$.
}
\end{figure}

To further investigate the exclusion of Type-I by the $Zt\bar{t}$ data, we present in \autoref{fig-BRH-Ztt} the branching ratios of $H$ as a function of $\mhh$ (left panel) and $\sg_{\rm NNLO}(gg \to H)\br(H \to ZA)\br(A \to t\bar{t})$ at the 13 TeV LHC as a function of $\mhh$ (right panel).\footnote{The results are nearly identical for Type-II.}
We set $\ma=365\gev$ and $\tb=1.28$, with $\mch=\mhh$ prohibiting the decay $H\to \ch\wpm$, 
as is the case for allowed parameter points even at Step (i). 
The results for Type-I and Type-II are indistinguishable.

Clearly, $\br(H \to ZA)$ is substantial because the $H$-$Z$-$A$ vertex is proportional to $\sba$. We also observe that $\br(H \to ZA)$ increases with $\mhh$, driven by two factors. First, for $\mhh \gg m_Z$, the Goldstone Boson Equivalence Theorem~\cite{Cornwall:1974km,Lee:1977eg,Chanowitz:1985hj,Gounaris:1986cr,Veltman:1989ud} applies, causing the longitudinal component of the $Z$ boson to behave like a Goldstone boson, with the amplitude for producing longitudinally polarized $Z$ bosons increasing with energy. Second, a larger $\mhh$ opens up more phase space for the decay.

In the right panel of \autoref{fig-BRH-Ztt}, we show $\sg_{\rm NNLO}(gg \to H)\br(H \to Z A)\br(A \to t\bar{t})$ (blue band) for $\tb = 1.28 \pm 0.128$. The gluon fusion cross section is calculated at next-to-next-to-leading order (NNLO) using the \textsc{SUSHI} 1.7.0 program~\cite{Harlander:2012pb,Harlander:2016hcx}, with branching ratios computed using the 2HDMC program. The red line indicates the observed 95\% upper limits from ATLAS~\cite{ATLAS:2023szc}, while the grey region is excluded by perturbativity constraints. It is clear that $H$ would need to be heavier than about 900 GeV to satisfy the ATLAS $Zt\bar{t}$ constraint. However, this mass range is already excluded by the perturbativity condition.

In conclusion, neither Type-I nor Type-II (and similarly, Type-X and Type-Y) can explain the CMS $t\bar{t}$ excess within the viable parameter space.

\section{Conclusions}
\label{sec:conclusions}

The intriguing $\ttop$ excess recently reported by the CMS Collaboration has generated excitement among those anticipating signs of new physics. This excess, observed in $\ttop$ production at an invariant mass of around 365 GeV, has prompted our investigation into whether it can be attributed to the pseudoscalar boson of conventional Two-Higgs-Doublet Models (2HDMs). This possibility is particularly compelling given that the angular distributions of charged leptons from top-pair decay strongly point to a pseudoscalar as the source of the enhancement.

While the toponium $\eta_t$ with $m_{\eta_t} \simeq 343$ GeV shows a marginally higher $-2 \ln L$ than the single pseudoscalar $A$ at 365 GeV, exploring both scenarios, along with other Beyond Standard Model (BSM) theories, remains valuable. Conventional 2HDMs, which naturally incorporate a pseudoscalar boson $A$, are among the first candidates one might consider. These models have been subjected to numerous theoretical and experimental constraints.

The best-fit parameters for the pseudoscalar boson are $m_A = 365$ GeV, $\Gamma_A/\ma =0.02$, and $\tan\beta = 1.28$. Accounting for experimental uncertainties, we allowed a 10\% variation in $\tan\beta$. With this consideration, we conducted a comprehensive scan of the allowed parameter space for Type-I and Type-II 2HDMs. Since the results of Type-I (II) are the same as Type-X (Y), our study effectively covers all four types. We systematically imposed constraints on the parameter space, including theoretical restrictions, electroweak precision data, perturbativity, Flavor-Changing Neutral Current (FCNC) constraints, and the most recent $t\bar t Z$ data.

Our analysis revealed that the perturbativity condition imposes relatively low upper bounds on $\mch$ and $\ma$ of approximately 723 GeV. Subsequently, upon applying FCNC constraints, we found that the entire parameter space for Type-II 2HDMs is ruled out, while only a small region survives for Type-I. However, even this remaining viable region in Type-I is ultimately eliminated when we consider the recent ATLAS $t \bar t Z$ data. This sequential application of constraints demonstrates the cumulative power of theoretical requirements and experimental observations in restricting the 2HDM parameter space.

The conclusion of our study is unequivocal: conventional 2HDMs (Types I, II, X, Y) lack viable parameter space to accommodate a pseudoscalar boson with mass $m_A =365$ GeV and $\tan\beta \simeq 1.3$, which is necessary to explain the observed excess in the $t\bar t$ production threshold region.

Nevertheless, this result does not preclude the possibility of explaining the excess through extensions or modifications to conventional 2HDMs. For instance, adding a singlet and a triple Higgs to the 2HDM, 
as proposed by Coloretti et al.~\cite{Coloretti:2023yyq}, 
could potentially account for the excess in the $t\bar t$ threshold region 
while also addressing other experimental anomalies.
A crucial test moving forward would be to scrutinize the threshold region and examine the interference pattern in the $m_{t\bar t}$ invariant-mass distribution. This analysis could enable the calculation of the phase between the Standard Model amplitude and the pseudoscalar amplitude, potentially pinpointing the nature of the BSM physics at play.

However, a note of caution is warranted. If the toponium effect is incorporated into the background fit, the contribution of a pseudoscalar boson diminishes, which allows larger values for $\tan\beta$. 
Then it is feasible to preserve some viable parameter space points even after imposing all the constraints considered in this work.

\acknowledgments
CTL is supported by the National Natural Science Foundation of China (NNSFC) under grant No.~12335005 and the Special funds for postdoctoral overseas recruitment, Ministry of Education of China.
K.C. is supported by the NSTC of Taiwan with grant number NSTC-113-2112-M-007-041-MY3.  
The work of DK, SL, and JS is supported by 
the National Research Foundation of Korea, Grant No.~NRF-2022R1A2C1007583. 

\bibliographystyle{JHEPMod}
\bibliography{A365-tt-2HDM}

\end{document}